\documentclass[aps,preprint,pra,longbibliography,showpacs]{revtex4-1}

\usepackage{graphicx}
\usepackage{dcolumn}
\usepackage{bm}

\usepackage{amssymb}
\usepackage{amsbsy}
\usepackage{color}

\providecommand\bnabla{\boldsymbol{\nabla}}
\providecommand\bcdot{\boldsymbol{\cdot}}

\newsavebox{\astrutbox}
\sbox{\astrutbox}{\rule[-5pt]{0pt}{20pt}}

\begin{document}

\preprint{APS/123-QED}

\title{Secondary instabilities in the flow past a cylinder: insights from a local stability analysis}

\author{Yogesh Jethani}
\author{Kamal Kumar}%
\author{A. Sameen}%
\author{Manikandan Mathur}%
 \email{manims@ae.iitm.ac.in}
\affiliation{%
Department of Aerospace Engineering, Indian Institute of Technology Madras, Chennai - 600036, India
}%


\date{\today}

\begin{abstract}
We perform a three-dimensional, short-wavelength stability analysis on the numerically simulated two-dimensional flow past a circular cylinder for Reynolds numbers in the range $50\le Re\le300$; here, $Re = U_{\infty}D/\nu$ with $U_\infty$, $D$ and $\nu$ being the free-stream velocity, the diameter of the cylinder and the kinematic viscosity of the fluid, respectively. For a given $Re$, inviscid local stability equations from the geometric optics approach are solved on three distinct closed fluid particle trajectories (denoted as orbits 1, 2 \& 3) for purely transverse perturbations. The inviscid instability on orbits 1 \& 2, which are symmetric counterparts of one another, is shown to undergo bifurcations at $Re\approx50$ and $Re\approx250$. Upon incorporating finite-wavenumber, finite-Reynolds number effects to compute corrected local instability growth rates, the inviscid instability on orbits 1 \& 2 is shown to be suppressed for $Re\lesssim262$. Orbits 1 \& 2 are thus shown to exhibit a synchronous instability for $Re\gtrsim262$, which is remarkably close to the critical Reynolds number for the mode-B secondary instability. Further evidence for the connection between the local instability on orbits 1 \& 2, and the mode-B secondary instability, is provided via a comparison of the growth rate variation with span-wise wavenumber between the local and global stability approaches. In summary, our results strongly suggest that the three-dimensional short-wavelength instability on orbits 1 \& 2 is a possible mechanism for the emergence of the mode B secondary instability.
\begin{description}
\item[Usage]
Secondary publications and information retrieval purposes.
\item[PACS numbers]
May be entered using the \verb+\pacs{#1}+ command.
\item[Structure]
You may use the \texttt{description} environment to structure your abstract;
use the optional argument of the \verb+\item+ command to give the category of each item. 
\end{description}
\end{abstract}

\pacs{Valid PACS appear here}
\maketitle


\section{Introduction} \label{sec:intro}
Various regimes occur in the flow past a circular cylinder as the Reynolds number $Re=U_\infty D/\nu$ is varied. Here, $U_\infty$ is the uniform free-stream velocity, $D$ the diameter of the cylinder and $\nu$ the kinematic viscosity of the fluid. For small $Re$ ($\lesssim 5$), the flow is steady and attached to the cylinder. At $Re\approx5$, flow separation occurs, resulting in the formation of a vortex pair in the near-wake steady separation bubble~\citep{zdravkovich1997}. At $Re\approx47$, a supercritical Hopf bifurcation leads to the time-periodic, laminar vortex shedding flow~\citep{provansal1987benard}. Laboratory experiments have shown two possible modes of vortex shedding, oblique and parallel, for $Re$ as low as $50$~\cite{williamson1989}. On further increasing $Re$, various secondary instabilities occur on the vortex shedding flow in the range $160\lesssim Re\lesssim260$~\citep{zhang1995transition,williamson1996vortex}. Linear stability analyses on appropriately chosen base flows have provided useful insights on the various transitions that occur in the flow past a circular cylinder~\citep{huerre1990local,williamson1996vortex}. For example, the absolute instability of the symmetric mode of perturbations in the time-averaged base flows at $Re\gtrsim47$ is attributed to the origin of self-sustained vortex shedding \citep{koch1985local,triantafyllou1986formation,oertel1990wakes}. In this study, we perform a three-dimensional, short-wavelength, local stability analysis in the two-dimensional near-wake region for $50\leq Re\leq300$, i.e. the vortex shedding regime.

Experimental and numerical studies \citep{williamson1996vortex} have revealed four different types of secondary instabilities that occur in the vortex shedding regime: the vortex adhesion mode \citep{williamson1992natural,zhang1995transition}, the two self-sustaining near-wake instabilities known as modes A \& B \citep{williamson1989,williamson1992natural,zhang1995transition}, and another near-wake instability mode C that occurs with suitable excitation \citep{zhang1995transition}. While all these instabilities can originate somewhere in the range of $160\lesssim Re\lesssim 230$, the actual state of the flow and the transition points depend significantly on various factors like end-conditions, external noise, and the path along which $Re$ is varied. The vortex dislocations associated with the vortex adhesion mode are observed in the range $160\lesssim Re\lesssim230$, with the number of adhesion points along the span of the cylinder increasing with $Re$. Mode A instability, characterized by its span-wise wavelength of around $3D$ to $4D$, is known to be dominant in the range $180\lesssim Re\lesssim 230$. Mode B, characterized by a span-wise wavelength of around $1D$, is first seen in laboratory experiments at $Re\approx230$, and subsequently becomes dominant at $Re\approx260$ \citep{thompson1996three}. Mode C, of characteristic span-wise wavelength of around $2D$, has been observed in the range $170<Re<270$ in experiments with a thin wire placed at specific locations around the cylinder. It is furthermore noteworthy that while modes A \& B are synchronous (same periodicity as the base flow), mode C is quasi-periodic \citep{zhang1995transition}.

A global numerical stability analysis on the two-dimensional periodic base flow revealed the emergence of an absolute linear instability at $Re \approx 188.5$ with a critical span-wise wavelength of around $3.96D$, and a second branch of linear instability at $Re\approx259$ with a critical span-wise wavelength of around $0.822D$~\citep{barkley1996three}. These results are qualitatively consistent with the mode A and mode B instabilities observed in experiments and numerical simulations.  To identify the mechanisms underlying the secondary instabilities, \citet{williamson1996vortex} and \citet{marxen2013vortex} investigated experimental and numerical data to conclude that mode A is associated with the elliptic instability of the vortex cores in the vortex street, and mode B with the braid regions between the shed vortices. In this study, we perform linear stability analysis on the fully non-parallel, unsteady, two-dimensional base flow, albeit with the restriction of short-wavelength perturbations.

We employ the local stability approach~\citep{lifschitz1991stabilityconditions}, which, based on the Wentzel-Kramers-Brillouin-Jeffreys (WKBJ) approximation~\citep{bender1999advanced}, investigates the evolution of three-dimensional, short-wavelength perturbations along fluid particle trajectories in a given base flow. This approach, which is computationally accessible even for strongly non-parallel flows, has been instrumental in understanding the mechanisms underlying elliptic instability~\citep{bayly1986elliptical,landman1987strainedvortices}, centrifugal instability~\citep{bayly1988centrifugal,nagarathinam2015} and hyperbolic instability~\citep{friedlander_vishik,leblanc1997single}. The local stability analysis has been particularly insightful in the studies on various idealized vortex models, including Stuart vortices \citep{godeferd2001zonal,mathur2014stuart} and Taylor-Green vortices~\citep{sipp_lauga_jacquin}. It has also been used to investigate numerically simulated base flows; for example, \citet{gallaire2007three} and \citet{citro_etal_2015} studied the flow over a bump and the incompressible open cavity flow, respectively, albeit only for transverse perturbation modes in the steady regime. Furthermore, \citet{giannetti2015wkbj} used this approach on the numerically simulated flow past a two-dimensional circular cylinder at $Re = 190$ and $260$.

In their study, \citet{giannetti2015wkbj} numerically solved the inviscid local stability equations on three different closed fluid particle trajectories for each of $Re = 190$ and 260. All three trajectories were found to be inviscidly unstable at $Re = 190$ \& $260$, with qualitatively similar Floquet exponents for corresponding trajectories at both the $Re$ values. They associate the instability on two of the trajectories with the asynchronous mode C instability, and the instability on the third trajectory with the synchronous modes A \& B. The relation between the inviscid local instabilities and the global mode stability analysis, however, remains unclear. In this paper, we perform local stability analysis in the entire range of $50\le Re\le300$ to investigate the dependence, with possible transitions, of the short-wavelength instabilities on $Re$. Finite $Re$, finite wavenumber corrections are inferentially incorporated, and comparisons between the local and global stability results are presented. We discover previously unknown bifurcations in the local instabilities, and discuss their potential relation with the evolution of secondary instabilities with $Re$.

The paper is organized as follows. In \S~\ref{sec:metho}, the details of the theory and the numerical construct are presented. This is followed up with the results and discussion of the local stability analysis on the fully unsteady flow in \S~\ref{sec:results_unsteady}. We present our conclusions in \S~\ref{sec:concl}.
\section{Methodology} \label{sec:metho}
We perform a local stability analysis on the incompressible, viscous, two-dimensional flow past a circular cylinder at various Reynolds numbers specified by $Re=50$, 60, 80, 100, 120, 140, 150, 160, 170, 180, 190, 200, 210, 220, 230, 240, 245, 248, 249, 249.5, 250, 250.5, 251, 252, 255, 260, 270, 280, 290 and 300, all of which are in the unsteady vortex shedding regime. Smaller steps in $Re$ around $Re\approx250$ ensured that the corresponding bifurcations were well-resolved.
\subsection{Base flow - numerical simulations}
The base flow is obtained by numerically solving the two-dimensional, incompressible mass and momentum equations for a uniform, horizontal free-stream incident on a circular cylinder on which the no-slip and no-normal-flow boundary conditions are satisfied. The numerical simulations were carried out in OpenFOAM \citep{weller1998tensorial} using a structured grid. The governing equations were solved using icoFoam, a finite-volume based incompressible, laminar Navier-Stokes solver that employs the PISO (Pressure Implicit with Splitting of Operator) algorithm. A rectangular computational domain of size 40$D$ $\times$ 20$D$ was used, with the inflow and the side boundaries at 10$D$ and the outflow boundary at $30D$  from the centre of the cylinder. At the inflow boundary, a uniform horizontal free-stream velocity $U_\infty$ is prescribed, whereas a uniform pressure boundary condition is used at the outflow boundary. At the two side boundaries, we assume no flow variations in a direction normal to them. The structured grid representing the computational domain was built such that the cylinder surface contained around 240 grid points, with a relatively coarser grid close to the domain boundaries. The grid size and the time step were appropriately chosen to ensure numerical convergence of the solutions at the largest $Re$, i.e. $Re = 300$. Additionally, we verified that our results agreed with the existing literature \citep{fey1998new,singh2005flow} on the values of Strouhal number and mean drag coefficients to within 3.5\% for all $Re$. In the rest of this paper, the spatial coordinates and the velocity field are non-dimensionalized by $D$ and $U_\infty$, respectively, with $D/U_\infty$ being the corresponding time scale to non-dimensionalize time. The cylinder orientation is such that the free-stream is in the direction of positive $x-$axis and the cylinder axis is along the $z-$axis, with $y-$axis perpendicular to both $x-$ and $z-$ axes.
\subsection{Local stability equations} \label{sec:LSA}
The local stability equations, i.e. the governing equations for the evolution of three-dimensional, short-wavelength, small perturbations are derived from the linearized mass and momentum equations. Within the WKBJ approximation, the velocity and pressure perturbations, $\boldsymbol{u}$ and $p$, respectively, are assumed to be of the following forms \citep{lifschitz1991stabilityconditions}:
\begin{equation}\label{eq:WKBJ_approx_u}
\boldsymbol{u}=\exp(i\phi(\boldsymbol{x},t)/\epsilon)[\boldsymbol{a}(\boldsymbol{x},t)+\epsilon\boldsymbol{a_1}(\boldsymbol{x},t)+...],
\end{equation}
\begin{equation}\label{eq:WKBJ_approx_p}
p=\exp(i\phi(\boldsymbol{x},t)/\epsilon)[\pi(\boldsymbol{x},t)+\epsilon\pi_1(\boldsymbol{x},t)+...],
\end{equation}
where $\boldsymbol{x}$ and $t$ denote space and time, respectively; $\phi(\boldsymbol{x},t)$ is any real scalar field and $\epsilon$ a small parameter. The complex amplitude of velocity perturbation at orders $\epsilon^0$ and $\epsilon^1$ are  $\boldsymbol{a}$ and ${\boldsymbol a_1}$, respectively; the corresponding pressure amplitudes are $\pi$ and $\pi_1$. The wave vector of the perturbations is given by $\boldsymbol{k} = \nabla\boldsymbol{\phi}/\epsilon$, highlighting that the analysis is restricted to short-wavelength perturbations. The inviscid local stability equations governing the evolution of $\boldsymbol{k}$ and $\boldsymbol{a}$ are~\citep{lifschitz1991stabilityconditions}:
\begin{equation}\label{eq:gov_eqn_for_k}
\frac{d\boldsymbol{k}}{dt}=-(\bnabla\boldsymbol{U_B})^T\bcdot\boldsymbol{k},
\end{equation}
\begin{equation}\label{eq:gov_eqn_for_a}
\frac{d\boldsymbol{a}}{dt}=-\bnabla\boldsymbol{U_B}\bcdot\boldsymbol{a}+\frac{2}{\mathopen|\boldsymbol{k}\mathclose|^2}[(\bnabla\boldsymbol{U_B}\bcdot\boldsymbol{a})\bcdot\boldsymbol{k}]\boldsymbol{k},
\end{equation}
where $d/dt = \partial/\partial t + \boldsymbol{U_B}\bcdot\boldsymbol{\nabla}$ is the total time derivative with respect to the base flow $\boldsymbol{U_B}=u_b\boldsymbol{\hat{e}_x}+v_b\boldsymbol{\hat{e}_y}$. Apart from the evolution equations \ref{eq:gov_eqn_for_k} \& \ref{eq:gov_eqn_for_a}, the continuity equation also requires $\boldsymbol{k}\bcdot\boldsymbol{a}=0$ to be satisfied. As noted in \citet{mathur2014stuart}, the solutions to equations \ref{eq:gov_eqn_for_k} \& \ref{eq:gov_eqn_for_a} always satisfy $\boldsymbol{k}\bcdot\boldsymbol{a}=0$ if the initial conditions do.

We restrict our studies to the solutions of the local stability equations (equations \ref{eq:gov_eqn_for_k}, \ref{eq:gov_eqn_for_a}) on closed trajectories in the base flow. The analysis is further restricted to those perturbations whose wave vector is periodic upon evolution (according to equation~\ref{eq:gov_eqn_for_k}) around one period of the closed trajectories, owing to which Floquet theory~\citep{chicone1999ordinary} is applicable to calculate the corresponding growth rates.
\subsection{Computation of growth rates} \label{sec:floq}
For all $Re$ in the range $50\leq Re \le 300$, the growth rates are computed for the three closed trajectories described in the beginning of $\S$~\ref{sec:results_unsteady} and shown in figure~\ref{fig:unsteady_flow_closed_traj}. The closed trajectories, with the time period the same as that of the base flow, are identified by calculating the fluid particle trajectories for a dense set of initial conditions in a sufficiently large domain in the near-wake of the cylinder.

The closed fluid particle trajectories are obtained by numerically integrating the equations $dx/dt=u_b(x,y,t)$ and $dy/dt=v_b(x,y,t)$ using the Runge-Kutta fourth-order scheme. Two-dimensional cubic interpolation in space is used to obtain the velocity field and its spatial derivatives on the trajectories, along with a linear interpolation in time. The time step of $0.05$ used in the numerical integration is small enough to ensure that a further reduction in the time step does not significantly alter the resulting growth rates, and causes no qualitative or quantitative changes to our conclusions. For the three closed trajectories in the unsteady flow for $50\leq Re\leq300$, we numerically verified that only purely transverse perturbations are periodic upon integration around the closed trajectories. Furthermore, it suffices to consider initial wave vectors of unit magnitude as the inviscid local stability equations~\ref{eq:gov_eqn_for_k} \& \ref{eq:gov_eqn_for_a} are linear in $\boldsymbol{k}$. Inviscid growth rates are therefore computed only for initial wave vectors specified by $\boldsymbol{k_i} = \boldsymbol{\hat{e}_z}$, for which $\boldsymbol{k}$ remains constant along any fluid particle trajectory (from equation~\ref{eq:gov_eqn_for_k}).

For a given closed trajectory and an initial wave vector, the governing equation~\ref{eq:gov_eqn_for_a} for the perturbation amplitude $\boldsymbol a$ is solved numerically using the Runge-Kutta fourth order scheme from $t=0$ to $t=T$, where $T$ is the time period of the flow. Solutions are obtained for three linearly independent initial conditions for $\boldsymbol{a}$: $\boldsymbol{a_{1,\,i}}=[1\quad 0\quad 0]$, $\boldsymbol{a_{2,\,i}}=[0\quad 1\quad 0]$ and $\boldsymbol{a_{3,\,i}}=[0\quad 0\quad 1]$, to obtain the final amplitude vectors at $t=T$ as $\boldsymbol{a_{1,\,f}}=[a_{x,1}\quad a_{y,1}\quad a_{z,1}]$, $\boldsymbol{a_{2,\,f}}=[a_{x,2}\quad a_{y,2}\quad a_{z,2}]$ and $\boldsymbol{a_{3,\,f}}=[a_{x,3}\quad a_{y,3}\quad a_{z,3}]$, respectively. The eigenvalues of the $3\times3$ matrix $M = \left[ a_{x,1} \; a_{x,2} \; a_{x,3};\; a_{y,1} \; a_{y,2} \; a_{y,3};\; a_{z,1} \; a_{z,2} \; a_{z,3} \right]$ represent the Floquet multipliers \citep{chicone1999ordinary}, denoted as $E_1$, $E_2$ and $E_3$. The corresponding Floquet exponents are $\sigma_j = (1/T)\log(E_j)$ ($j=1,2,3$), with the inviscid growth rate given by $\sigma_0^{re}= \max\left(\Re\left(\left\lbrace \sigma_1,\sigma_2,\sigma_3\right\rbrace\right)\right)$, where $\Re$ denotes the real part. The imaginary part, $\sigma_0^{im}$ of the complex growth rate is given by $\sigma_0^{im}=\Im(\log E_m)/T$, where $E_m$ is the eigenvalue that corresponds to the inviscid growth rate $\sigma_0^{re}$ and $\Im$ denotes the imaginary part. In summary, $(1/T)\log(E_m)=\sigma_0^{re}+i\sigma_0^{im}$. The methodology we adopt to incorporate the effects of finite $Re$ and finite wavenumber are discussed in section~\ref{sec:viscous_growth_rates}.
\section{Results and discussion}\label{sec:results_unsteady}
\begin{figure}
\begin{center}
\includegraphics[width=0.84\textwidth,angle=0]{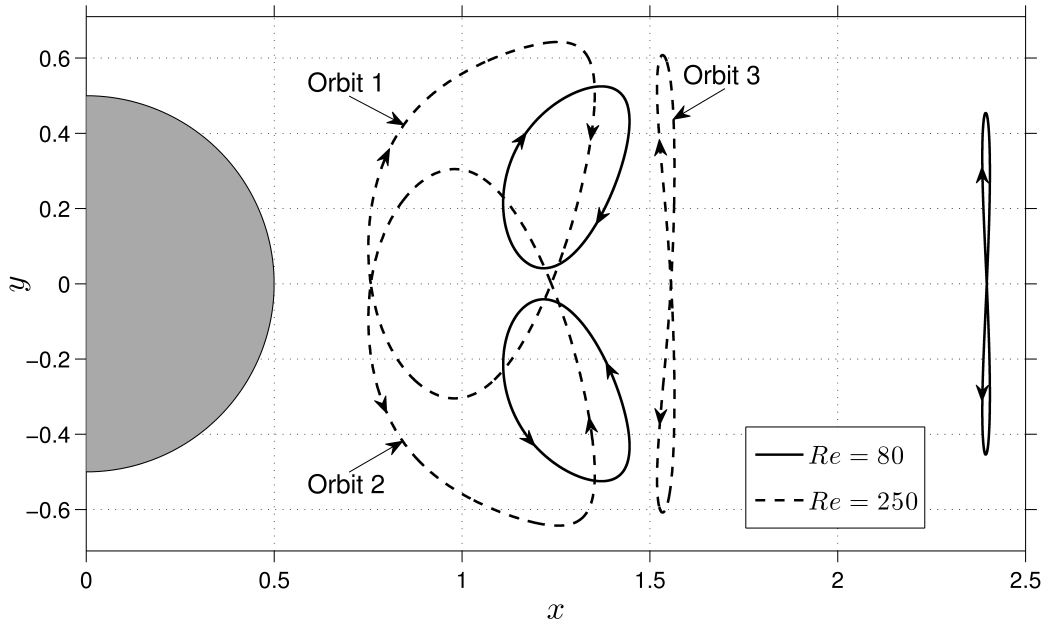}
\caption{The closed trajectories in the unsteady flow for $Re = 80$ (solid lines) and $250$ (dashed lines). Orbits 1, 2 \& 3 are explicitly labelled for $Re = 250$. Topologically similar three closed trajectories are found for all $Re\in\left[50,300\right]$. The uniform free-stream veloctiy $U_\infty$ is along the positive $x-$axis.\label{fig:unsteady_flow_closed_traj}}
\end{center}
\end{figure}
For every $Re\in[50,300]$, we find three distinct closed fluid particle trajectories whose time period $T$ is the same as that of the flow. In figure~\ref{fig:unsteady_flow_closed_traj}, we show the three trajectories for $Re = 250$ (dashed lines), which are labelled as orbits 1, 2 and 3. Topologically similar orbits for $Re = 80$ are also shown in figure~\ref{fig:unsteady_flow_closed_traj} (solid lines). Orbits 1 \& 2, which are symmetric counterparts of each other but with a time lag of $T/2$, are non-self-intersecting, whereas orbit 3 intersects itself on the centreline. With increasing $Re$, the three orbits move closer to the cylinder and simultaneously increase in their spatial extent. Previous studies by \citet{giannetti2015wkbj} identified the same three closed orbits, but only for $Re = 190$ \& $260$. As mentioned in section~\ref{sec:floq}, for all the three orbits at a given $Re$, only purely transverse wave vectors are periodic with the time period $T$, owing to which we compute growth rates only for $\boldsymbol{k_i}  = \boldsymbol{\hat{e}_z}$. Furthermore, orbits 1 \& 2 correspond to the same growth rates, and hence we plot all results only for orbits 1 \& 3.
\subsection{Inviscid growth rates}\label{sec:inviscid_growth_rates}
\begin{figure}
\begin{center}
\includegraphics[width=1\textwidth,angle=0]{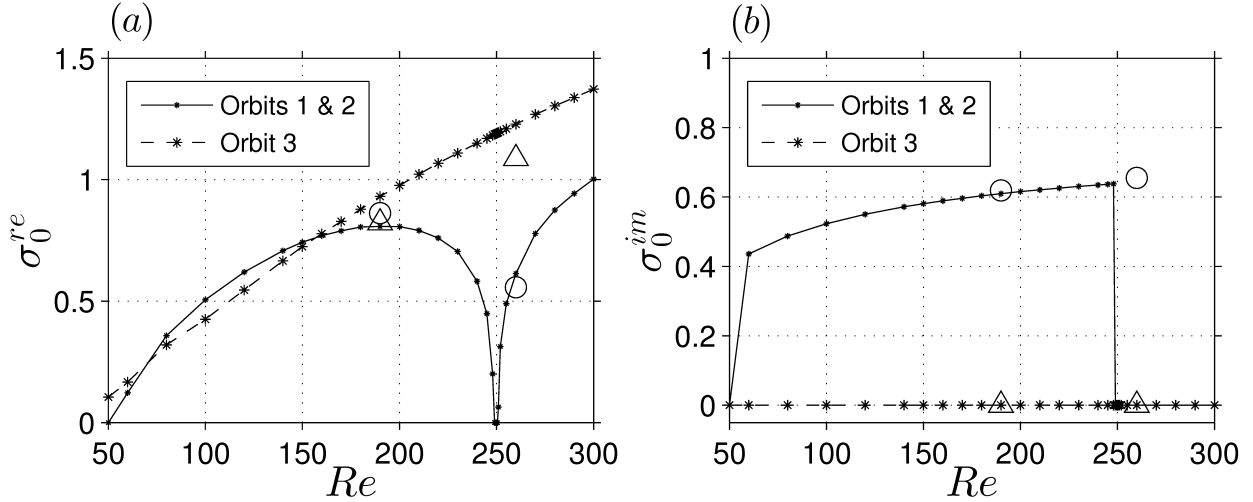}
\caption{$(a)$ Inviscid growth rate $\sigma_0^{re}$ and $(b)$ the imaginary part $\sigma_0^{im}$ of the complex growth rate, for orbits 1 \& 2 (solid line) and orbit 3 (dashed line) plotted as a function of $Re$ for purely transverse perturbations. Corresponding values from \citet{giannetti2015wkbj} for $Re=190$ \& $260$ are indicated by {\scriptsize$\bigcirc$} (orbits 1 \& 2) and {\normalsize$\triangle$} (orbit 3). The markers on the curves indicate the actual values of $Re$ at which the growth rate calculations were performed. \label{fig:unsteady_flow_growth_rates}}
\end{center}
\end{figure}

In figure~\ref{fig:unsteady_flow_growth_rates}$(a)$, we plot the inviscid growth rate $\sigma_0^{re}$ for orbit 1 (solid line) and orbit 3 (dashed line) as a function of $Re$. For orbit 1, $\sigma_0^{re}$ is zero at $Re = 50$, which is immediately followed by a bifurcation as indicated by the positive $\sigma_0^{re}$ at $Re = 60$. For $Re\lesssim190$, $\sigma_0^{re}$ increases monotonically, attaining a local maximum of $\sigma_0^{re} \approx 0.8$ at $Re \approx 190$, before decreasing to zero at $Re \approx 250$. Remarkably, orbit 1 is stable for a small range of $Re$ around $Re = 250$, above which it again becomes unstable with $\sigma_0^{re}$ increasing to 1 at $Re = 300$. The two bifurcations associated with orbit 1, immediately below and above $Re = 250$, are discussed in more detail later in this section. In contrast, orbit 3 is unstable in the entire range of $\left[50,300\right]$, with its $\sigma_0^{re}$ monotonically increasing with $Re$. The instability associated with orbit 1 slightly dominates over that of orbit 3 in the range $70\lesssim Re\lesssim157$, above which the instability of orbit 3 clearly dominates up to $Re = 300$. Also included in figure~\ref{fig:unsteady_flow_growth_rates}$(a)$ are the growth rate estimates of \citet{giannetti2015wkbj} at $Re = 190$ \& $260$, which are in reasonable agreement with our results. The difference between our computed growth rates and those of \citet{giannetti2015wkbj} may be attributed to differences in the respective numerical schemes used to generate the base flows.

We recall here that the imaginary part of the complex growth rate is given by $\sigma_0^{im}=\Im(\log E_m)/T$, where $E_m$ is the eigenvalue that corresponds to the inviscid growth rate. In figure~\ref{fig:unsteady_flow_growth_rates}$(b)$, we plot $\sigma_0^{im}$ as a function of $Re$ for orbits 1 \& 3. For orbit 1, $\sigma_0^{re}>0$ and $\sigma_0^{im}=\pi/T$ ($T$ is the time period of the closed trajectory) in the range $60\le Re\le248$, indicating that the corresponding eigenvalue $E_m$ is real and less than -1. For $251\le Re\le300$, $\sigma_0^{re}>0$ with $\sigma_0^{im}  = 0$, i.e. the corresponding eigenvalue $E_m$ is real and greater than 1. The instability on orbit 1 is therefore seen to switch from being asynchronous ($\sigma_0^{im}\ne 0$) for $Re\le248$ to synchronous ($\sigma_0^{im} = 0$) for $Re\ge251$. In physical terms, a synchronous instability has the same time period as that of the base flow. In contrast, at $Re = 260$, \citet{giannetti2015wkbj} report $\sigma_0^{im}=\pi/T$, suggesting that $E_m$ is real and less than -1. We recall, however, from figure~\ref{fig:unsteady_flow_growth_rates}$(a)$ that $\sigma_0^{re}$ for orbit 1 from our calculations is in good quantitative agreement with that of \citet{giannetti2015wkbj} at $Re = 260$. We are unable to identify the source of the discrepancy in $\sigma_0^{im}$ for orbit 1 as \citet{giannetti2015wkbj} report their results only at $Re = 190$ \& $Re = 260$. For orbit 3, the eigenvalue $E_m$ is real and positive at all $Re$,  resulting in $\sigma_0^{im}$ being zero at all $Re$. 

To investigate the bifurcations in the stability on orbit 1 at $Re\approx50$ and $Re\approx250$, we track the evolution of the non-trivial Floquet exponents $\sigma_1$ and $\sigma_2$ with $Re$; the trivial Floquet exponent is $\sigma_3 = 0$. $\sigma_1$ and $\sigma_2$ are defined such that $\Re(\sigma_1)\ge \Re(\sigma_2)$. At $Re =50$, $\sigma_1 = \sigma_2^*$ ($^*$ denotes complex conjugate) with $\Re(\sigma_1)<0$ and $\Re(\sigma_2)<0$, whereas at $Re=60$, $\sigma_1 = -\sigma_2^*$ with $\Im(\sigma_1)=\Im(\sigma_2) = \pi/T$. In other words, the  stability characteristics on orbit 1 switch from being stable-focus-like at $Re = 50$ to unstable-saddle-like at $Re=60$, but with non-zero imaginary parts in the Floquet exponents. The stability property on orbit 1 remains unstable-saddle-like, with $\Im(\sigma_1)\ne 0$ and $\Im(\sigma_2)\ne 0$, for $Re<249$. 

\begin{figure}
\begin{center}
\includegraphics[width=0.9\textwidth,angle=0]{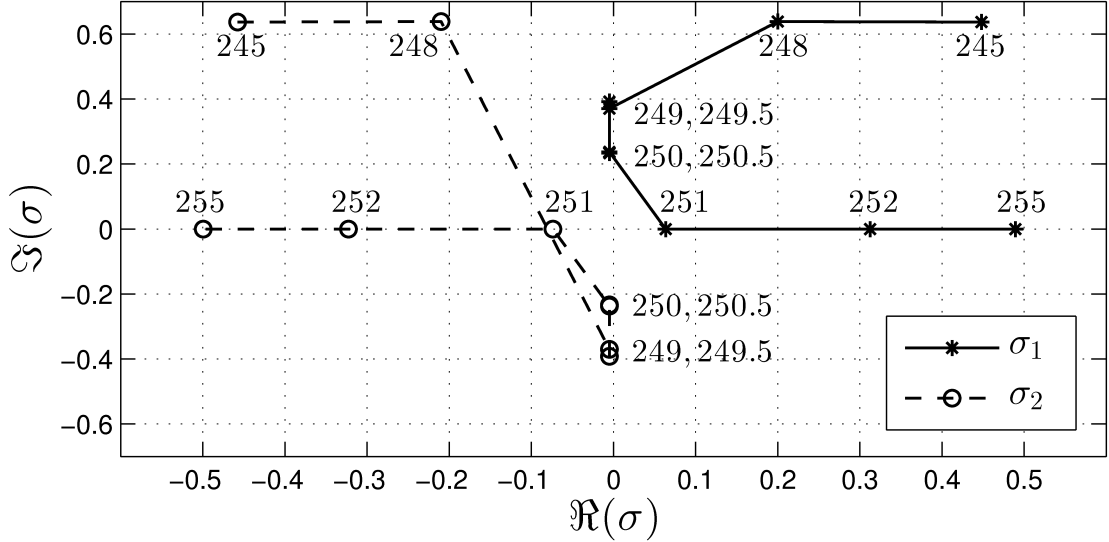}
\caption{Path traced by the non-trivial Floquet exponents $\sigma_1$ (solid line) and $\sigma_2$ (dashed line) on the complex plane as $Re$ is varied in the vicinity of $Re=250$, with $\sigma_1$ and $\sigma_2$ defined such that $\Re(\sigma_1)\ge\Re(\sigma_2)$. Indicated right next to every datapoint is the corresponding value of $Re$.\label{fig:eigenvalue_variation_around_250}}
\end{center}
\end{figure}

For the two bifurcations at around $Re = 250$, we plot $\sigma_1$ and $\sigma_2$ on the complex plane as $Re$ is varied in the neighbourhood of $Re = 250$ (figure~\ref{fig:eigenvalue_variation_around_250}).  For $Re = 245$ and 248, $\Re(\sigma_1)>0$ and $\Re(\sigma_2)<0$, with both the exponents having the same positive imaginary part. The exponents then switch to being complex conjugates with negative real parts at $Re  = 249$. This  bifurcation from unstable-saddle-like behaviour (with $\Im(\sigma_1)\ne 0$ and $\Im(\sigma_2)\ne 0$) to stable-focus-like behaviour is the reverse of what occurs at $Re\approx50$. The Floquet exponents remain as complex conjugates with $\Re(\sigma_1)<0$ and $\Re(\sigma_2)<0$ for $Re = 249.5$, 250 \& 250.5. We then observe a switch to Floquet exponents with $\sigma_1 = -\sigma_2^*$ at $Re = 251$, with $\Im(\sigma_1)=\Im(\sigma_2)=0$. This bifurcation is therefore from stable-focus-like to unstable-saddle-like properties. For all $Re\ge 251$, the instability property on orbit 1 remains saddle-like with $\sigma_1 = -\sigma_2^*$ and $\Im(\sigma_1)=\Im(\sigma_2)=0$.

\subsection{Viscous growth rates}\label{sec:viscous_growth_rates}
To explore the relation between our results and existing knowledge on secondary instabilities in the cylinder wake, we incorporate finite-wavenumber, finite-$Re$ effects in the following manner~\citep{gallaire2007three}
\begin{equation}\label{eq:gr_corr}
\sigma_{\nu}(\beta,Re) = \sigma_0(Re) - \frac{\beta^2}{Re} - \frac{A(Re)}{\beta},
\end{equation}
where $\sigma_{\nu}$ is the corrected growth rate, $\sigma_0$ the inviscid growth rate from the local stability calculations, $\beta$ the transverse wave number, and $A$ a model parameter. The first correction term, $\beta^2/Re$, in equation~\ref{eq:gr_corr} follows from the study of~\citet{landman1987strainedvortices} that showed that weak viscous effects always serve to suppress the inviscid local instabilities. The second correction term, $A/\beta$, in equation~\ref{eq:gr_corr} is based on a previous study~\citep{bayly1988centrifugal} that proposed the construction of localized eigenmodes from local stability calculations. The results of \citet{bayly1988centrifugal} were specific to centrifugal instability on a streamline with locally maximum inviscid growth rate in a steady flow, implying that the validity of equation~\ref{eq:gr_corr} for a closed trajectory in an unsteady flow is unknown. However, equation~\ref{eq:gr_corr} has been employed for centrifugal and non-centrifugal-type instabilities, which don't necessarily satisfy all the assumptions of \citet{bayly1988centrifugal}, with reasonable accuracy~\citep{sipp1999vortices,gallaire2007three}. This suggests that equation~\ref{eq:gr_corr} may represent a reasonably accurate generic model for finite-$\beta$, finite-$Re$ corrections.

The model parameter, $A$ in equation~\ref{eq:gr_corr} is assumed to be independent of $\beta$ \citep{gallaire2007three} and a function of the base flow Reynolds number $Re$ only. To estimate $A(Re)$, we use inputs from known secondary instability characteristics. The self-sustained mode-B secondary instability has a characteristic span-wise wavelength of around 1$D$, making it more likely to be captured in a short-wavelength framework than mode-A (characteristic span-wise wavelength $\approx 4D$). Furthermore, global mode Floquet analysis reveals the emergence of an instability at $Re \approx 260$ with a dominant mode of span-wise wavelength around $0.822D$ \citep{barkley1996three}, and is associated with the mode-B instability observed in experiments and three-dimensional numerical simulations. Therefore, we explore the possible connection between our local instability calculations and the mode-B secondary instability, specifically focusing on the relevance of the bifurcation in the instability of orbit 1 at $Re\approx250$.

\begin{figure}
\begin{center}
\includegraphics[width=0.9\textwidth,angle=0]{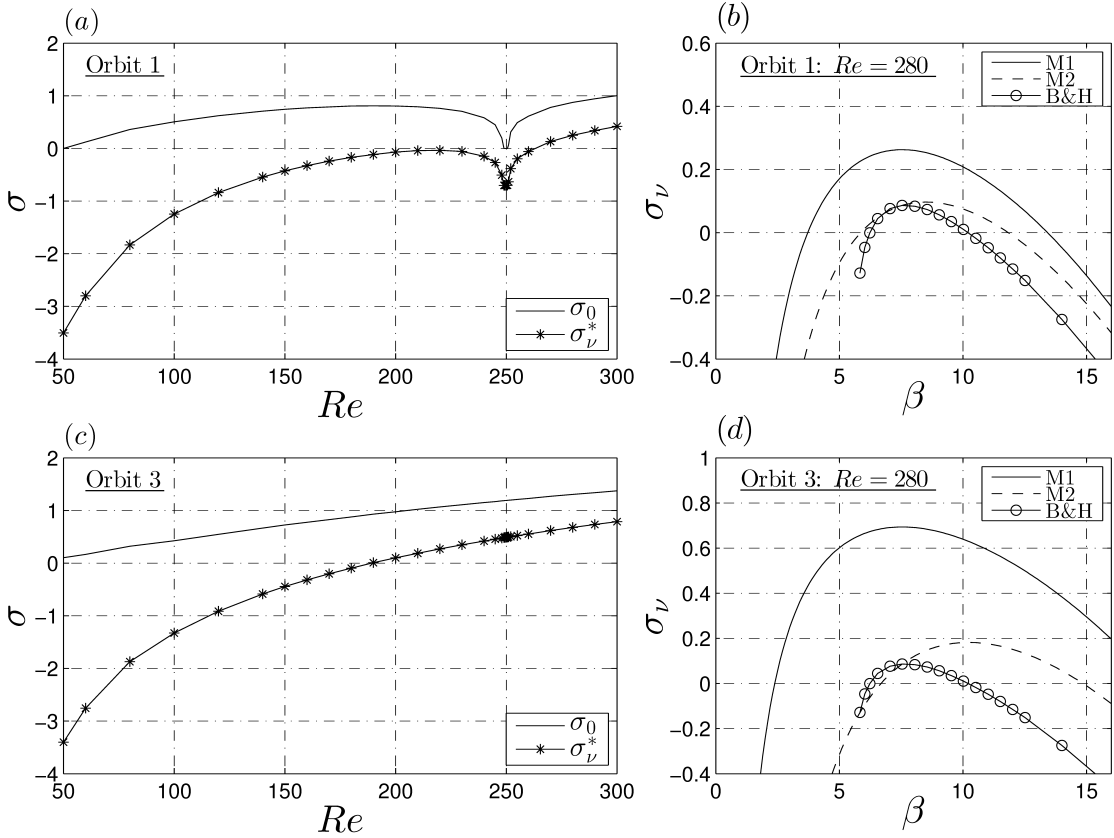}
\caption{$(a)$ The maximum growth rate $\sigma_{\nu}^\ast=\sigma_0-3\beta^{\ast 2}/Re$ (with $\beta^\ast = 7.64$) as a function of $Re$ (solid line with markers) for orbits 1 and 2. The inviscid growth rate $\sigma_0$ for orbit 1 is reproduced here for comparison (solid line with no markers). $(b)$ The corrected growth rate $\sigma_\nu$ (equation~\ref{eq:gr_corr}) as a function of $\beta$ using two methodologies M1 and M2 for orbit 1 at $Re=280$; for comparison, the global stability growth rates from~\citet{barkley1996three} are plotted using open circles. $(c)$ and $(d)$: Similar data as in $(a)$ and $(b)$, respectively is plotted for oribt 3. \label{fig:corr_growth_rates}}
\end{center}
\end{figure}
\citet{barkley1996three} report that the most unstable perturbation mode at the birth of the new instability at $Re \approx 260$ corresponds to a span-wise wavenumber of $\beta = \beta^* \approx 7.64$. While the exact values of $\beta^\ast$ are not reported for $Re> 260$ in \citet{barkley1996three}, the most unstable wavenumber is known to be around the same value as that for $Re = 260$. For $\beta = \beta^*$ to be the most unstable transverse wavenumber in the local stability analysis, i.e. for $\sigma_\nu$ (equation \ref{eq:gr_corr}) to attain a maximum at $\beta = \beta^*$, we require
\begin{equation}\label{eq:beta_max_law}
A=\frac{2\beta^{*3}}{Re},
\end{equation}
thus providing us with an estimate for $A$. The resulting maximum growth rate $\sigma_\nu(\beta^*,Re)  = \sigma_0(Re) - 3\beta^{\ast 2}/Re$ is denoted as $\sigma_\nu^*$. In figure~\ref{fig:corr_growth_rates}$(a)$, we plot $\sigma_\nu^*$ (solid line with markers) as a function of $Re$ for orbit 1. Strikingly, based on $\sigma_\nu^*$, orbit 1 is stable for all $Re\lesssim262$. This critical $Re$ of 262 above which orbit 1 is unstable is close to the mode-B critical $Re$ of 260 reported by~\citet{barkley1996three}. Additionally, as noted in figure~\ref{fig:unsteady_flow_growth_rates}$(b)$, the instability on orbit 1 is synchronous ($\sigma_0^{im}=0$) for $Re\ge 262$, and hence leading us to suggest that it is indeed closely related to the synchronous mode-B secondary instability.

To further explore the relation between the local instability on orbit 1 and the mode-B secondary instability, we evaluate the extent to which equation~\ref{eq:gr_corr} captures the growth rate variation with $\beta$. In figure~\ref{fig:corr_growth_rates}$(b)$, we plot $\sigma_\nu$ for orbit 1 as a function of $\beta$ at $Re = 280$ (solid line), with the value of $A$ chosen based on equation~\ref{eq:beta_max_law}; such a choice for $A$ is denoted as methodology M1. The solid line, representing methodology M1, in figure~\ref{fig:corr_growth_rates}$(b)$ is in reasonable qualitative agreement with the growth rate variation based on the results of \citet{barkley1996three} (solid line with markers). Specifically, the range of unstable $\beta$ based on $\sigma_\nu$ is $[3.72, 13.42]$, whereas the unstable range reported by \citet{barkley1996three} is $[6.23,10.2]$. In figure~\ref{fig:corr_growth_rates}$(b)$, we also plot $\sigma_\nu$ vs. $\beta$ using an alternate method M2 to estimate $A$, where we match $\sigma_\nu(\beta^*)$ with the corresponding value of growth rate at $\beta^\ast$ from \citet{barkley1996three}. While methodology M2 does not guarantee that $\sigma_\nu$ attains a maximum at $\beta = \beta^*$, we find remarkable qualitative and quantitative agreement with \citet{barkley1996three}. Based on methodology M2, the range of unstable $\beta$ is $[5.87, 11.84]$. In summary, the results from figures~\ref{fig:corr_growth_rates}$(a)$ \& $(b)$ indicate that the mode-B secondary instability is possibly a manifestation of the local instability on orbits 1 \& 2. 

To investigate the effects of finite-$\beta$, finite-$Re$ corrections on the local instability identified on orbit 3, we plot $\sigma_\nu^*$ as a function of $Re$ in figure~\ref{fig:corr_growth_rates}$(c)$, with $A$ again estimated using $\beta^* = 7.64$. Based on the corrected growth rate, orbit 3 is stable for all $Re$ less than the critical value of $Re\approx 190$, a value of Reynolds number that lies in the so-called transition regime \citep{williamson1996vortex}. However, $Re = 190$ represents the critical Reynolds number of the mode-A secondary instability, whose characteristic transverse wavelength is $4D$. Therefore, it is not clear what the implications of the transition at $Re = 190$ in figure~\ref{fig:corr_growth_rates}$(c)$ are. We also note that orbit 3 is of noticeably smaller spatial extent than that of orbit 1, and also contains segments where its radius of curvature is negligibly small, potentially raising questions on whether a span-wise wavelength of $1D$ can be considered ``short-wavelength" for orbit 3. In a manner similar to figure~\ref{fig:corr_growth_rates}$(b)$, we plot $\sigma_\nu$ as a function of $\beta$ (for both methodologies M1 and M2) at $Re=280$ for orbit 3, along with the mode-B instability growth rate variation reported by \citet{barkley1996three}. For both methodologies M1 and M2, a significantly larger range of beta is unstable when compared with the estimate from global analysis. This further suggests that the local instability on orbit 3, probably, has no relation to the mode-B secondary instability.

Finally, we performed similar calculations as in figure~\ref{fig:corr_growth_rates} to explore the relation between the local instabilities and the mode-A secondary instability. The span-wise wavelength associated with the mode-A secondary instability is around $4D$ ($\beta \approx 1.59$), and hence not necessarily a ``short-wavelength instability". Assuming a $\beta^*$ of 1.59, finite-$Re$, finite-$\beta$ corrections do not significantly modify the inviscid growth rates on orbits 1 or 3. Therefore, our study is inconclusive about the role of local instabilities in the mode-A secondary instability.

The related study by \citet{giannetti2015wkbj} performed similar inviscid local stability calculations on the three closed trajectories for $Re = 190$ \& 260. The inviscid instability on orbit 1 was reported to be asynchronous for both $Re$, leading \citet{giannetti2015wkbj} to suggest that it is related to the mode-C secondary instability. Orbit 3 was found to display a synchronous instability, and was linked with both modes A \& B secondary instabilities. Our study, however, finds that the inviscid instability on orbit 1 undergoes an asynchronous-to-synchronous bifurcation at $Re\approx 250$. Additionally, with finite-wavenumber, finite-$Re$ corrections, the synchronous instability on orbit 1 is shown to occur for $Re\gtrsim262$ only. Furthermore, we also find the mode B span-wise wavelength of around $1D$ to be consistent with the local instability on orbit 1, but not on orbit 3.

\section{Conclusions}\label{sec:concl}
In this paper, we have presented a local stability analysis in the near-wake region resulting from the uniform flow past a two-dimensional circular cylinder for Reynolds numbers in the range $50\le Re\le300$. The inviscid local stability equations were solved on closed fluid particle trajectories in the unsteady flow for purely transverse perturbations. Three closed trajectories with the time period the same as that of the base flow were identified for all $Re\in\left[50,300\right]$. Two of these closed trajectories, denoted orbits 1 \& 2, are non-self-intersecting and symmetric counterparts of each other. The third closed trajectory, referred to as orbit 3, is self-intersecting. The inviscid growth rate associated with orbits 1 \& 2 undergoes bifurcations at $Re\approx 50$ and $Re\approx 250$, with the instabilities in the ranges $50\lesssim Re \lesssim248$ and $251\lesssim Re\le300$ being asynchronous and synchronous, respectively.  In contrast, orbit 3 was found to be inviscidly unstable in the entire range of $\left[50,300\right]$, with the corresponding growth rate increasing monotonically with $Re$.

Finite-wavenumber, finite-Reynolds number corrections on the computed inviscid growth rates were then obtained by assuming that the most unstable perturbation mode occurs for the mode-B span-wise wavelength of $0.822D$. Based on this corrected growth rate, the inviscid instability on orbits 1 \& 2 is suppressed for $Re\lesssim262$, and a synchronous instability occurs for $Re>262$. This transition Reynolds number of $Re\approx262$ is remarkably close to the mode B critical Reynolds number from global stability analysis, experiments and numerical simulations. Additionally, the corrected growth rate variation with the span-wise wavenumber for $Re=280$ shows excellent qualitative and quantitative agreement with the corresponding mode B instability growth rates from the global analysis. These results strongly suggest that the three-dimensional, short-wavelength instability on orbit 1 is closely connected to the mode B secondary instability in the wake of a circular cylinder. The physical relevance of the inviscid local instability on orbit 3 is unclear.

In the future, it may be worth designing direct numerical simulations with the aim of studying the evolution of localized three-dimensional perturbations on the closed trajectories in the two-dimensional cylinder wake, which may lead to devising flow control strategies. Also, it would be interesting to explore the existence of other closed trajectories in the time-periodic cylinder wake, whose time period may be any integer multiple of the flow time period.

We thank S. Ajith Kumar for his help with the numerical simulations of the base flow in the early phase of this study.

\bibliography{paperPRF_refs}

\begin{thebibliography}{32}%
\makeatletter
\providecommand \@ifxundefined [1]{%
 \@ifx{#1\undefined}
}%
\providecommand \@ifnum [1]{%
 \ifnum #1\expandafter \@firstoftwo
 \else \expandafter \@secondoftwo
 \fi
}%
\providecommand \@ifx [1]{%
 \ifx #1\expandafter \@firstoftwo
 \else \expandafter \@secondoftwo
 \fi
}%
\providecommand \natexlab [1]{#1}%
\providecommand \enquote  [1]{``#1''}%
\providecommand \bibnamefont  [1]{#1}%
\providecommand \bibfnamefont [1]{#1}%
\providecommand \citenamefont [1]{#1}%
\providecommand \href@noop [0]{\@secondoftwo}%
\providecommand \href [0]{\begingroup \@sanitize@url \@href}%
\providecommand \@href[1]{\@@startlink{#1}\@@href}%
\providecommand \@@href[1]{\endgroup#1\@@endlink}%
\providecommand \@sanitize@url [0]{\catcode `\\12\catcode `\$12\catcode
  `\&12\catcode `\#12\catcode `\^12\catcode `\_12\catcode `\%12\relax}%
\providecommand \@@startlink[1]{}%
\providecommand \@@endlink[0]{}%
\providecommand \url  [0]{\begingroup\@sanitize@url \@url }%
\providecommand \@url [1]{\endgroup\@href {#1}{\urlprefix }}%
\providecommand \urlprefix  [0]{URL }%
\providecommand \Eprint [0]{\href }%
\providecommand \doibase [0]{http://dx.doi.org/}%
\providecommand \selectlanguage [0]{\@gobble}%
\providecommand \bibinfo  [0]{\@secondoftwo}%
\providecommand \bibfield  [0]{\@secondoftwo}%
\providecommand \translation [1]{[#1]}%
\providecommand \BibitemOpen [0]{}%
\providecommand \bibitemStop [0]{}%
\providecommand \bibitemNoStop [0]{.\EOS\space}%
\providecommand \EOS [0]{\spacefactor3000\relax}%
\providecommand \BibitemShut  [1]{\csname bibitem#1\endcsname}%
\let\auto@bib@innerbib\@empty
\bibitem [{\citenamefont {Zdravkovich}(1997)}]{zdravkovich1997}%
  \BibitemOpen
  \bibfield  {author} {\bibinfo {author} {\bibfnamefont {M.~M.}\ \bibnamefont
  {Zdravkovich}},\ }\href {http://books.google.co.in/books?id=w8tSQwAACAAJ}
  {\emph {\bibinfo {title} {`Flow Around Circular Cylinders: Volume I:
  Fundamentals'}}},\ Flow Around Circular Cylinders: A Comprehensive Guide
  Through Flow Phenomena, Experiments, Applications, Mathematical Models, and
  Computer Simulations\ (\bibinfo  {publisher} {OUP Oxford},\ \bibinfo {year}
  {1997})\BibitemShut {NoStop}%
\bibitem [{\citenamefont {Provansal}\ \emph {et~al.}(1987)\citenamefont
  {Provansal}, \citenamefont {Mathis},\ and\ \citenamefont
  {Boyer}}]{provansal1987benard}%
  \BibitemOpen
  \bibfield  {author} {\bibinfo {author} {\bibfnamefont {M.}~\bibnamefont
  {Provansal}}, \bibinfo {author} {\bibfnamefont {C.}~\bibnamefont {Mathis}}, \
  and\ \bibinfo {author} {\bibfnamefont {L.}~\bibnamefont {Boyer}},\ }\bibfield
   {title} {\enquote {\bibinfo {title} {B{\'e}nard-von {K\'a}rm{\'a}n
  instability: transient and forced regimes},}\ }\href@noop {} {\bibfield
  {journal} {\bibinfo  {journal} {J. Fluid Mech.}\ }\textbf {\bibinfo {volume}
  {182}},\ \bibinfo {pages} {1--22} (\bibinfo {year} {1987})}\BibitemShut
  {NoStop}%
\bibitem [{\citenamefont {Williamson}(1989)}]{williamson1989}%
  \BibitemOpen
  \bibfield  {author} {\bibinfo {author} {\bibfnamefont {C.~H.~K.}\
  \bibnamefont {Williamson}},\ }\bibfield  {title} {\enquote {\bibinfo {title}
  {Oblique and parallel modes of vortex shedding in the wake of a circular
  cylinder at low {R}eynolds numbers},}\ }\href@noop {} {\bibfield  {journal}
  {\bibinfo  {journal} {J. Fluid Mech.}\ }\textbf {\bibinfo {volume} {206}},\
  \bibinfo {pages} {579--627} (\bibinfo {year} {1989})}\BibitemShut {NoStop}%
\bibitem [{\citenamefont {Zhang}\ \emph {et~al.}(1995)\citenamefont {Zhang},
  \citenamefont {Fey}, \citenamefont {Noack}, \citenamefont {K{\"o}nig},\ and\
  \citenamefont {Eckelmann}}]{zhang1995transition}%
  \BibitemOpen
  \bibfield  {author} {\bibinfo {author} {\bibfnamefont {H.-Q.}\ \bibnamefont
  {Zhang}}, \bibinfo {author} {\bibfnamefont {U.}~\bibnamefont {Fey}}, \bibinfo
  {author} {\bibfnamefont {B.~R.}\ \bibnamefont {Noack}}, \bibinfo {author}
  {\bibfnamefont {M.}~\bibnamefont {K{\"o}nig}}, \ and\ \bibinfo {author}
  {\bibfnamefont {H.}~\bibnamefont {Eckelmann}},\ }\bibfield  {title} {\enquote
  {\bibinfo {title} {On the transition of the cylinder wake},}\ }\href@noop {}
  {\bibfield  {journal} {\bibinfo  {journal} {Phys. Fluids}\ }\textbf {\bibinfo
  {volume} {7}},\ \bibinfo {pages} {779--794} (\bibinfo {year}
  {1995})}\BibitemShut {NoStop}%
\bibitem [{\citenamefont {Williamson}(1996)}]{williamson1996vortex}%
  \BibitemOpen
  \bibfield  {author} {\bibinfo {author} {\bibfnamefont {C.~H.~K.}\
  \bibnamefont {Williamson}},\ }\bibfield  {title} {\enquote {\bibinfo {title}
  {Vortex dynamics in the cylinder wake},}\ }\href@noop {} {\bibfield
  {journal} {\bibinfo  {journal} {Annu. Rev. Fluid Mech.}\ }\textbf {\bibinfo
  {volume} {28}},\ \bibinfo {pages} {477--539} (\bibinfo {year}
  {1996})}\BibitemShut {NoStop}%
\bibitem [{\citenamefont {Huerre}\ and\ \citenamefont
  {Monkewitz}(1990)}]{huerre1990local}%
  \BibitemOpen
  \bibfield  {author} {\bibinfo {author} {\bibfnamefont {P.}~\bibnamefont
  {Huerre}}\ and\ \bibinfo {author} {\bibfnamefont {P.~A.}\ \bibnamefont
  {Monkewitz}},\ }\bibfield  {title} {\enquote {\bibinfo {title} {Local and
  global instabilities in spatially developing flows},}\ }\href@noop {}
  {\bibfield  {journal} {\bibinfo  {journal} {Annu. Rev. Fluid Mech.}\ }\textbf
  {\bibinfo {volume} {22}},\ \bibinfo {pages} {473--537} (\bibinfo {year}
  {1990})}\BibitemShut {NoStop}%
\bibitem [{\citenamefont {Koch}(1985)}]{koch1985local}%
  \BibitemOpen
  \bibfield  {author} {\bibinfo {author} {\bibfnamefont {W.}~\bibnamefont
  {Koch}},\ }\bibfield  {title} {\enquote {\bibinfo {title} {Local instability
  characteristics and frequency determination of self-excited wake flows},}\
  }\href@noop {} {\bibfield  {journal} {\bibinfo  {journal} {J. Sound and
  Vib.}\ }\textbf {\bibinfo {volume} {99}},\ \bibinfo {pages} {53--83}
  (\bibinfo {year} {1985})}\BibitemShut {NoStop}%
\bibitem [{\citenamefont {Triantafyllou}\ \emph {et~al.}(1986)\citenamefont
  {Triantafyllou}, \citenamefont {Triantafyllou},\ and\ \citenamefont
  {Chryssostomidis}}]{triantafyllou1986formation}%
  \BibitemOpen
  \bibfield  {author} {\bibinfo {author} {\bibfnamefont {G.~S.}\ \bibnamefont
  {Triantafyllou}}, \bibinfo {author} {\bibfnamefont {M.~S.}\ \bibnamefont
  {Triantafyllou}}, \ and\ \bibinfo {author} {\bibfnamefont {C.}~\bibnamefont
  {Chryssostomidis}},\ }\bibfield  {title} {\enquote {\bibinfo {title} {On the
  formation of vortex streets behind stationary cylinders},}\ }\href@noop {}
  {\bibfield  {journal} {\bibinfo  {journal} {J. Fluid Mech.}\ }\textbf
  {\bibinfo {volume} {170}},\ \bibinfo {pages} {461--477} (\bibinfo {year}
  {1986})}\BibitemShut {NoStop}%
\bibitem [{\citenamefont {Oertel}(1990)}]{oertel1990wakes}%
  \BibitemOpen
  \bibfield  {author} {\bibinfo {author} {\bibfnamefont {{Jr, H.}}\
  \bibnamefont {Oertel}},\ }\bibfield  {title} {\enquote {\bibinfo {title}
  {Wakes behind blunt bodies},}\ }\href@noop {} {\bibfield  {journal} {\bibinfo
   {journal} {Annu. Rev. Fluid Mech.}\ }\textbf {\bibinfo {volume} {22}},\
  \bibinfo {pages} {539--562} (\bibinfo {year} {1990})}\BibitemShut {NoStop}%
\bibitem [{\citenamefont {Williamson}(1992)}]{williamson1992natural}%
  \BibitemOpen
  \bibfield  {author} {\bibinfo {author} {\bibfnamefont {C.~H.~K.}\
  \bibnamefont {Williamson}},\ }\bibfield  {title} {\enquote {\bibinfo {title}
  {The natural and forced formation of spot-like `vortex dislocations' in the
  transition of a wake},}\ }\href@noop {} {\bibfield  {journal} {\bibinfo
  {journal} {J. Fluid Mech.}\ }\textbf {\bibinfo {volume} {243}},\ \bibinfo
  {pages} {393--441} (\bibinfo {year} {1992})}\BibitemShut {NoStop}%
\bibitem [{\citenamefont {Thompson}\ \emph {et~al.}(1996)\citenamefont
  {Thompson}, \citenamefont {Hourigan},\ and\ \citenamefont
  {Sheridan}}]{thompson1996three}%
  \BibitemOpen
  \bibfield  {author} {\bibinfo {author} {\bibfnamefont {M.}~\bibnamefont
  {Thompson}}, \bibinfo {author} {\bibfnamefont {K.}~\bibnamefont {Hourigan}},
  \ and\ \bibinfo {author} {\bibfnamefont {J.}~\bibnamefont {Sheridan}},\
  }\bibfield  {title} {\enquote {\bibinfo {title} {Three-dimensional
  instabilities in the wake of a circular cylinder},}\ }\href@noop {}
  {\bibfield  {journal} {\bibinfo  {journal} {Exp.Thermal Fluid Sci.}\ }\textbf
  {\bibinfo {volume} {12}},\ \bibinfo {pages} {190--196} (\bibinfo {year}
  {1996})}\BibitemShut {NoStop}%
\bibitem [{\citenamefont {Barkley}\ and\ \citenamefont
  {Henderson}(1996)}]{barkley1996three}%
  \BibitemOpen
  \bibfield  {author} {\bibinfo {author} {\bibfnamefont {D.}~\bibnamefont
  {Barkley}}\ and\ \bibinfo {author} {\bibfnamefont {R.~D.}\ \bibnamefont
  {Henderson}},\ }\bibfield  {title} {\enquote {\bibinfo {title}
  {Three-dimensional {F}loquet stability analysis of the wake of a circular
  cylinder},}\ }\href@noop {} {\bibfield  {journal} {\bibinfo  {journal} {J.
  Fluid Mech.}\ }\textbf {\bibinfo {volume} {322}},\ \bibinfo {pages}
  {215--241} (\bibinfo {year} {1996})}\BibitemShut {NoStop}%
\bibitem [{\citenamefont {Marxen}\ \emph {et~al.}(2013)\citenamefont {Marxen},
  \citenamefont {Lang},\ and\ \citenamefont {Rist}}]{marxen2013vortex}%
  \BibitemOpen
  \bibfield  {author} {\bibinfo {author} {\bibfnamefont {O.}~\bibnamefont
  {Marxen}}, \bibinfo {author} {\bibfnamefont {M.}~\bibnamefont {Lang}}, \ and\
  \bibinfo {author} {\bibfnamefont {U.}~\bibnamefont {Rist}},\ }\bibfield
  {title} {\enquote {\bibinfo {title} {Vortex formation and vortex breakup in a
  laminar separation bubble},}\ }\href@noop {} {\bibfield  {journal} {\bibinfo
  {journal} {J. Fluid Mech.}\ }\textbf {\bibinfo {volume} {728}},\ \bibinfo
  {pages} {58--90} (\bibinfo {year} {2013})}\BibitemShut {NoStop}%
\bibitem [{\citenamefont {Lifschitz}\ and\ \citenamefont
  {Hameiri}(1991)}]{lifschitz1991stabilityconditions}%
  \BibitemOpen
  \bibfield  {author} {\bibinfo {author} {\bibfnamefont {A.}~\bibnamefont
  {Lifschitz}}\ and\ \bibinfo {author} {\bibfnamefont {E.}~\bibnamefont
  {Hameiri}},\ }\bibfield  {title} {\enquote {\bibinfo {title} {Local stability
  conditions in fluid dynamics},}\ }\href {\doibase
  http://dx.doi.org/10.1063/1.858153} {\bibfield  {journal} {\bibinfo
  {journal} {Phys. Fluids A}\ }\textbf {\bibinfo {volume} {3}},\ \bibinfo
  {pages} {2644--2651} (\bibinfo {year} {1991})}\BibitemShut {NoStop}%
\bibitem [{\citenamefont {Bender}\ and\ \citenamefont
  {Orszag}(1999)}]{bender1999advanced}%
  \BibitemOpen
  \bibfield  {author} {\bibinfo {author} {\bibfnamefont {C.~M.}\ \bibnamefont
  {Bender}}\ and\ \bibinfo {author} {\bibfnamefont {S.~A.}\ \bibnamefont
  {Orszag}},\ }\href@noop {} {\emph {\bibinfo {title} {Advanced Mathematical
  Methods for Scientists and Engineers - Asymptotic Methods and Perturbation
  Theory}}}\ (\bibinfo  {publisher} {Springer},\ \bibinfo {year}
  {1999})\BibitemShut {NoStop}%
\bibitem [{\citenamefont {Bayly}(1986)}]{bayly1986elliptical}%
  \BibitemOpen
  \bibfield  {author} {\bibinfo {author} {\bibfnamefont {B.~J.}\ \bibnamefont
  {Bayly}},\ }\bibfield  {title} {\enquote {\bibinfo {title} {Three-dimensional
  instability of elliptical flow},}\ }\href
  {http://link.aps.org/doi/10.1103/PhysRevLett.57.2160} {\bibfield  {journal}
  {\bibinfo  {journal} {Phys. Rev. Lett.}\ }\textbf {\bibinfo {volume} {57
  (17)}},\ \bibinfo {pages} {2160--2163} (\bibinfo {year} {1986})}\BibitemShut
  {NoStop}%
\bibitem [{\citenamefont {Landman}\ and\ \citenamefont
  {Saffman}(1987)}]{landman1987strainedvortices}%
  \BibitemOpen
  \bibfield  {author} {\bibinfo {author} {\bibfnamefont {M.~J.}\ \bibnamefont
  {Landman}}\ and\ \bibinfo {author} {\bibfnamefont {P.~G.}\ \bibnamefont
  {Saffman}},\ }\bibfield  {title} {\enquote {\bibinfo {title} {The
  three-dimensional instability of strained vortices in a viscous fluid},}\
  }\href {\doibase http://dx.doi.org/10.1063/1.866124} {\bibfield  {journal}
  {\bibinfo  {journal} {Phys. Fluids}\ }\textbf {\bibinfo {volume} {30}},\
  \bibinfo {pages} {2339--2342} (\bibinfo {year} {1987})}\BibitemShut {NoStop}%
\bibitem [{\citenamefont {Bayly}(1988)}]{bayly1988centrifugal}%
  \BibitemOpen
  \bibfield  {author} {\bibinfo {author} {\bibfnamefont {B.~J.}\ \bibnamefont
  {Bayly}},\ }\bibfield  {title} {\enquote {\bibinfo {title} {Three-dimensional
  centrifugal-type instabilities in inviscid two-dimensional flows},}\ }\href
  {\doibase http://dx.doi.org/10.1063/1.867002} {\bibfield  {journal} {\bibinfo
   {journal} {Phys. Fluids}\ }\textbf {\bibinfo {volume} {31}},\ \bibinfo
  {pages} {56--64} (\bibinfo {year} {1988})}\BibitemShut {NoStop}%
\bibitem [{\citenamefont {Nagarathinam}\ \emph {et~al.}(2015)\citenamefont
  {Nagarathinam}, \citenamefont {Sameen},\ and\ \citenamefont
  {Mathur}}]{nagarathinam2015}%
  \BibitemOpen
  \bibfield  {author} {\bibinfo {author} {\bibfnamefont {D.}~\bibnamefont
  {Nagarathinam}}, \bibinfo {author} {\bibfnamefont {A.}~\bibnamefont
  {Sameen}}, \ and\ \bibinfo {author} {\bibfnamefont {M.}~\bibnamefont
  {Mathur}},\ }\bibfield  {title} {\enquote {\bibinfo {title} {Centrifugal
  instability in nonaxisymmetric vortices},}\ }\href@noop {} {\bibfield
  {journal} {\bibinfo  {journal} {J. Fluid Mech.}\ }\textbf {\bibinfo {volume}
  {769}},\ \bibinfo {pages} {26--45} (\bibinfo {year} {2015})}\BibitemShut
  {NoStop}%
\bibitem [{\citenamefont {Friedlander}\ and\ \citenamefont
  {Vishik}(1991)}]{friedlander_vishik}%
  \BibitemOpen
  \bibfield  {author} {\bibinfo {author} {\bibfnamefont {S.}~\bibnamefont
  {Friedlander}}\ and\ \bibinfo {author} {\bibfnamefont {M.~M.}\ \bibnamefont
  {Vishik}},\ }\bibfield  {title} {\enquote {\bibinfo {title} {Instability
  criteria for the flow of an inviscid incompressible fluid},}\ }\href@noop {}
  {\bibfield  {journal} {\bibinfo  {journal} {Phys. Rev. Lett.}\ }\textbf
  {\bibinfo {volume} {66}},\ \bibinfo {pages} {2204--2206} (\bibinfo {year}
  {1991})}\BibitemShut {NoStop}%
\bibitem [{\citenamefont {Leblanc}(1997)}]{leblanc1997single}%
  \BibitemOpen
  \bibfield  {author} {\bibinfo {author} {\bibfnamefont {S.}~\bibnamefont
  {Leblanc}},\ }\bibfield  {title} {\enquote {\bibinfo {title} {Stability of
  stagnation points in rotating flows},}\ }\href@noop {} {\bibfield  {journal}
  {\bibinfo  {journal} {Phys. Fluids}\ }\textbf {\bibinfo {volume} {9}},\
  \bibinfo {pages} {3566--3569} (\bibinfo {year} {1997})}\BibitemShut {NoStop}%
\bibitem [{\citenamefont {Godeferd}\ \emph {et~al.}(2001)\citenamefont
  {Godeferd}, \citenamefont {Cambon},\ and\ \citenamefont
  {Leblanc}}]{godeferd2001zonal}%
  \BibitemOpen
  \bibfield  {author} {\bibinfo {author} {\bibfnamefont {F.~S.}\ \bibnamefont
  {Godeferd}}, \bibinfo {author} {\bibfnamefont {C.}~\bibnamefont {Cambon}}, \
  and\ \bibinfo {author} {\bibfnamefont {S.}~\bibnamefont {Leblanc}},\
  }\bibfield  {title} {\enquote {\bibinfo {title} {Zonal approach to
  centrifugal, elliptic and hyperbolic instabilities in {S}tuart vortices with
  external rotation},}\ }\href@noop {} {\bibfield  {journal} {\bibinfo
  {journal} {J. Fluid Mech.}\ }\textbf {\bibinfo {volume} {449}},\ \bibinfo
  {pages} {1--37} (\bibinfo {year} {2001})}\BibitemShut {NoStop}%
\bibitem [{\citenamefont {Mathur}\ \emph {et~al.}(2014)\citenamefont {Mathur},
  \citenamefont {Ortiz}, \citenamefont {Dubos},\ and\ \citenamefont
  {Chomaz}}]{mathur2014stuart}%
  \BibitemOpen
  \bibfield  {author} {\bibinfo {author} {\bibfnamefont {M.}~\bibnamefont
  {Mathur}}, \bibinfo {author} {\bibfnamefont {S.}~\bibnamefont {Ortiz}},
  \bibinfo {author} {\bibfnamefont {T.}~\bibnamefont {Dubos}}, \ and\ \bibinfo
  {author} {\bibfnamefont {J.~M.}\ \bibnamefont {Chomaz}},\ }\bibfield  {title}
  {\enquote {\bibinfo {title} {Effects of an axial flow on the centrifugal,
  elliptic and hyperbolic instabilities in {S}tuart vortices},}\ }\href@noop {}
  {\bibfield  {journal} {\bibinfo  {journal} {J. Fluid Mech.}\ }\textbf
  {\bibinfo {volume} {758}},\ \bibinfo {pages} {565--585} (\bibinfo {year}
  {2014})}\BibitemShut {NoStop}%
\bibitem [{\citenamefont {Sipp}\ \emph
  {et~al.}(1999{\natexlab{a}})\citenamefont {Sipp}, \citenamefont {Lauga},\
  and\ \citenamefont {Jacquin}}]{sipp_lauga_jacquin}%
  \BibitemOpen
  \bibfield  {author} {\bibinfo {author} {\bibfnamefont {D.}~\bibnamefont
  {Sipp}}, \bibinfo {author} {\bibfnamefont {E.}~\bibnamefont {Lauga}}, \ and\
  \bibinfo {author} {\bibfnamefont {L.}~\bibnamefont {Jacquin}},\ }\bibfield
  {title} {\enquote {\bibinfo {title} {Vortices in rotating systems:
  {C}entrifugal, elliptic and hyperbolic type instabilities},}\ }\href@noop {}
  {\bibfield  {journal} {\bibinfo  {journal} {Phys. Fluids}\ }\textbf {\bibinfo
  {volume} {11}},\ \bibinfo {pages} {3716--3728} (\bibinfo {year}
  {1999}{\natexlab{a}})}\BibitemShut {NoStop}%
\bibitem [{\citenamefont {Gallaire}\ \emph {et~al.}(2007)\citenamefont
  {Gallaire}, \citenamefont {Marquillie},\ and\ \citenamefont
  {Ehrenstein}}]{gallaire2007three}%
  \BibitemOpen
  \bibfield  {author} {\bibinfo {author} {\bibfnamefont {F.}~\bibnamefont
  {Gallaire}}, \bibinfo {author} {\bibfnamefont {M.}~\bibnamefont
  {Marquillie}}, \ and\ \bibinfo {author} {\bibfnamefont {U.}~\bibnamefont
  {Ehrenstein}},\ }\bibfield  {title} {\enquote {\bibinfo {title}
  {Three-dimensional transverse instabilities in detached boundary layers},}\
  }\href@noop {} {\bibfield  {journal} {\bibinfo  {journal} {J. Fluid Mech.}\
  }\textbf {\bibinfo {volume} {571}},\ \bibinfo {pages} {221--233} (\bibinfo
  {year} {2007})}\BibitemShut {NoStop}%
\bibitem [{\citenamefont {Citro}\ \emph {et~al.}(2015)\citenamefont {Citro},
  \citenamefont {Giannetti}, \citenamefont {Brandt},\ and\ \citenamefont
  {Luchini}}]{citro_etal_2015}%
  \BibitemOpen
  \bibfield  {author} {\bibinfo {author} {\bibfnamefont {V.}~\bibnamefont
  {Citro}}, \bibinfo {author} {\bibfnamefont {F.}~\bibnamefont {Giannetti}},
  \bibinfo {author} {\bibfnamefont {L.}~\bibnamefont {Brandt}}, \ and\ \bibinfo
  {author} {\bibfnamefont {P.}~\bibnamefont {Luchini}},\ }\bibfield  {title}
  {\enquote {\bibinfo {title} {Linear three-dimensional global and asymptotic
  stability analysis of incompressible open cavity flow},}\ }\href@noop {}
  {\bibfield  {journal} {\bibinfo  {journal} {J. Fluid Mech.}\ }\textbf
  {\bibinfo {volume} {768}},\ \bibinfo {pages} {113--140} (\bibinfo {year}
  {2015})}\BibitemShut {NoStop}%
\bibitem [{\citenamefont {Giannetti}(2015)}]{giannetti2015wkbj}%
  \BibitemOpen
  \bibfield  {author} {\bibinfo {author} {\bibfnamefont {F.}~\bibnamefont
  {Giannetti}},\ }\bibfield  {title} {\enquote {\bibinfo {title} {{WKBJ}
  analysis in the periodic wake of a cylinder},}\ }\href@noop {} {\bibfield
  {journal} {\bibinfo  {journal} {Theoretical and Applied Mechanics Letters}\
  }\textbf {\bibinfo {volume} {5}},\ \bibinfo {pages} {107--110} (\bibinfo
  {year} {2015})}\BibitemShut {NoStop}%
\bibitem [{\citenamefont {Weller}\ \emph {et~al.}(1998)\citenamefont {Weller},
  \citenamefont {Tabor}, \citenamefont {Jasak},\ and\ \citenamefont
  {Fureby}}]{weller1998tensorial}%
  \BibitemOpen
  \bibfield  {author} {\bibinfo {author} {\bibfnamefont {H.~G.}\ \bibnamefont
  {Weller}}, \bibinfo {author} {\bibfnamefont {G.}~\bibnamefont {Tabor}},
  \bibinfo {author} {\bibfnamefont {H.}~\bibnamefont {Jasak}}, \ and\ \bibinfo
  {author} {\bibfnamefont {C.}~\bibnamefont {Fureby}},\ }\bibfield  {title}
  {\enquote {\bibinfo {title} {A tensorial approach to computational continuum
  mechanics using object-oriented techniques},}\ }\href@noop {} {\bibfield
  {journal} {\bibinfo  {journal} {Computers in physics}\ }\textbf {\bibinfo
  {volume} {12}},\ \bibinfo {pages} {620--631} (\bibinfo {year}
  {1998})}\BibitemShut {NoStop}%
\bibitem [{\citenamefont {Fey}\ \emph {et~al.}(1998)\citenamefont {Fey},
  \citenamefont {K{\"o}nig},\ and\ \citenamefont {Eckelmann}}]{fey1998new}%
  \BibitemOpen
  \bibfield  {author} {\bibinfo {author} {\bibfnamefont {U.}~\bibnamefont
  {Fey}}, \bibinfo {author} {\bibfnamefont {M.}~\bibnamefont {K{\"o}nig}}, \
  and\ \bibinfo {author} {\bibfnamefont {H.}~\bibnamefont {Eckelmann}},\
  }\bibfield  {title} {\enquote {\bibinfo {title} {A new
  strouhal--reynolds-number relationship for the circular cylinder in the range
  47},}\ }\href@noop {} {\bibfield  {journal} {\bibinfo  {journal} {Physics of
  Fluids}\ }\textbf {\bibinfo {volume} {10}},\ \bibinfo {pages} {1547--1549}
  (\bibinfo {year} {1998})}\BibitemShut {NoStop}%
\bibitem [{\citenamefont {Singh}\ and\ \citenamefont
  {Mittal}(2005)}]{singh2005flow}%
  \BibitemOpen
  \bibfield  {author} {\bibinfo {author} {\bibfnamefont {S.~P.}\ \bibnamefont
  {Singh}}\ and\ \bibinfo {author} {\bibfnamefont {S.}~\bibnamefont {Mittal}},\
  }\bibfield  {title} {\enquote {\bibinfo {title} {Flow past a cylinder: shear
  layer instability and drag crisis},}\ }\href@noop {} {\bibfield  {journal}
  {\bibinfo  {journal} {International Journal for Numerical Methods in Fluids}\
  }\textbf {\bibinfo {volume} {47}},\ \bibinfo {pages} {75--98} (\bibinfo
  {year} {2005})}\BibitemShut {NoStop}%
\bibitem [{\citenamefont {Chicone}(1999)}]{chicone1999ordinary}%
  \BibitemOpen
  \bibfield  {author} {\bibinfo {author} {\bibfnamefont {C.~C.}\ \bibnamefont
  {Chicone}},\ }\href@noop {} {\emph {\bibinfo {title} {Ordinary differential
  equations with applications}}}\ (\bibinfo  {publisher} {Springer},\ \bibinfo
  {year} {1999})\BibitemShut {NoStop}%
\bibitem [{\citenamefont {Sipp}\ \emph
  {et~al.}(1999{\natexlab{b}})\citenamefont {Sipp}, \citenamefont {Lauga},\
  and\ \citenamefont {Jacquin}}]{sipp1999vortices}%
  \BibitemOpen
  \bibfield  {author} {\bibinfo {author} {\bibfnamefont {Denis}\ \bibnamefont
  {Sipp}}, \bibinfo {author} {\bibfnamefont {E}~\bibnamefont {Lauga}}, \ and\
  \bibinfo {author} {\bibfnamefont {L}~\bibnamefont {Jacquin}},\ }\bibfield
  {title} {\enquote {\bibinfo {title} {Vortices in rotating systems:
  Centrifugal, elliptic and hyperbolic type instabilities},}\ }\href@noop {}
  {\bibfield  {journal} {\bibinfo  {journal} {Physics of Fluids}\ }\textbf
  {\bibinfo {volume} {11}},\ \bibinfo {pages} {3716--3728} (\bibinfo {year}
  {1999}{\natexlab{b}})}\BibitemShut {NoStop}%
\end{thebibliography}%
\end{document}